\documentclass[conference,twocolumn]{IEEEtran}

\usepackage{graphicx}
\usepackage{color}
\usepackage{placeins}
\usepackage{float}
\usepackage{tabularx,colortbl}
\usepackage{amssymb}
\usepackage{amsthm}
\usepackage{cite}
\usepackage{amsmath}
\usepackage{epstopdf}
\usepackage{stfloats}
\theoremstyle{plain}

\theoremstyle{plain}

\theoremstyle{plain}
\newcommand{\tabincell}[2]{\begin{tabular}{@{}#1@{}}#2\end{tabular}}

\IEEEoverridecommandlockouts

\begin{document}
%
\title{Deep Learning-Based Power Control for Uplink Cell-Free Massive MIMO Systems}
\author{Yongshun~Zhang, Jiayi~Zhang, Yu~Jin, Stefano~Buzzi, Bo~Ai
\thanks{
This work was supported in part by Fundamental Research Funds for the Central Universities (2020JBZD005), National Natural Science Foundation of China (61971027, U1834210, and 61961130391), in part by Beijing Natural Science Foundation (L202013).}
\thanks{Y. Zhang, J. Zhang and Y. Jin are with School of Electronic and Information Engineering, Beijing Jiaotong University, Beijing 100044, China. They are also with Frontiers Science Center for Smart High-speed Railway System, Beijing Jiaotong University, Beijing 100044, China.}
\thanks{S. Buzzi is with Department of Electrical and Information Engineering, University of Cassino and Lazio Meridionale, Cassino, Italy.}
\thanks{B. Ai is with State Key Laboratory of Rail Traffic Control and Safety, Beijing Jiaotong University, Beijing 100044, China.}
}

\maketitle

\begin{abstract}
In this paper, a general framework for deep learning-based power control methods for max-min, max-product and max-sum-rate optimization in uplink cell-free massive multiple-input multiple-output (CF mMIMO) systems is proposed. Instead of using supervised learning, the proposed method relies on unsupervised learning, in which optimal power allocations are not required to be known, and thus has low training complexity. More specifically, a deep neural network (DNN) is trained to learn the map between fading coefficients and power coefficients within short time and with low computational complexity. It is interesting to note that the spectral efficiency of CF mMIMO systems with the proposed method outperforms previous optimization methods for max-min optimization and fits well for both max-sum-rate and max-product optimizations.
\end{abstract}


\IEEEpeerreviewmaketitle

\section{Introduction}
By using a large number of access points (APs) to serve a small number of user equipments (UEs), CF mMIMO can provide uniform coverage and large spectral efficiency (SE) \cite{ngo2017cell,zhang2020prospective,zhang2019cell}. There are no cells or boundaries in CF mMIMO, and all APs are connected with a central processing unit (CPU). The uplink SE of four implementations with spatially correlated fading and arbitrary linear processing has been analyzed in \cite{bjornson2019making}. A new framework for precoding and combing methods for uplink/downlink signal processing has been instead proposed by exploiting the dynamic cooperation cluster concept \cite{bjornson2020scalable, chen2020structured}, which was proved to be scalable. The performance of a CF mMIMO system with multi-antenna APs and phase shift was studied in \cite{wang2020uplink, ozdogan2019performance}, with spatially correlated Rician fading channels taken into consideration. Besides, the performance of unmanned aerial vehicle communications and channel aging in CF mMIMO were considered in \cite{zheng2021uav, zheng2021impact}.

One of the main challenges in uplink CF mMIMO systems is the optimal power control of UEs. Due to the rapid changes of small-scale fading, it is difficult to optimize power control coefficients. In mMIMO systems, however, due to channel hardening, we can optimize the transmission power based on only large-scale fading (LSF) coefficients and spatial correlation, instead of bringing small-scale fading coefficients into consideration \cite{ghazanfari2020enhanced}. To provide uniform service to all UEs, the max-min power control was proposed for CF mMIMO systems in \cite{ngo2017cell}. Furthermore, the authors proposed to maximize the geometric mean of the per-cell max-min SE to provide more fairness towards weak users \cite{ghazanfari2020enhanced}. A scalable method named fractional power control for uplink power control and downlink power allocation in CF mMIMO systems, relying only on LSF coefficients has been analyzed in \cite{nikbakht2020uplink}. To reduce the effect of pilot contamination, a pilot power control method that minimizes the mean-squared error of the channel estimation by choosing the pilot power control coefficients was proposed in \cite{mai2018pilot}. However, these non-convex power control methods require high computational complexity and their performance needs further improvement.

As one of the most efficient machine learning methods, deep learning (DL) has attracted much interest for the solution of optimal power control strategies in cellular and CF mMIMO systems. Compared with traditional optimization methods (e.g., greedy-like and convex optimization algorithms), deep learning techniques can hugely improve SE and greatly reduce computational complexity. Inspired by this important merit, recent works have used DL/unsupervised learning to solve the challenging power allocation problem in CF mMIMO systems. For example, a network was proposed to learn the map between the positions of users and the max-min or max-product optimal power control policies in the cellular mMIMO systems \cite{sanguinetti2018deep}. A DNN was proposed for max-min or max-sum-rate optimal power control policies to obtain optimal power control coefficients in CF mMIMO systems \cite{d2019uplink}. The authors in \cite{nikbakht2019unsupervised} trained a feedforward neural network (NN) based on unsupervised learning in uplink CF mMIMO systems to obtain the map between LSF coefficients and power control policies (e.g., soft max-min and max-prod). The advantage of the proposed method in \cite{nikbakht2019unsupervised} is that only the LSF coefficients are used as inputs. A feedforward NN was trained for uplink and downlink C-RAN systems to achieve optimal power allocation \cite{nikbakht2020unsupervised}. Moreover, a DL/unsupervised learning-based power control algorithm was proposed to solve the max-min optimization problem in CF mMIMO systems \cite{rajapaksha2021deep}, which also considered online learning. The author in \cite{zhao2020power} proposed a DNN to learn the optimal power coefficients in a time-division duplex based sub-6GHz network, taking the long-term fading information as input.

However, most of the aforementioned works are based on supervised learning, where networks consume a large number of training epochs and datasets. To tackle this problem, in this paper, we develop a DL-based power control method as a robust and low-complexity technique to solve the power control problem in CF mMIMO. The contributions of this paper are as follows:

\begin{itemize}
\item The proposed methods rely only on LSF coefficients without the location information of UEs. LSF coefficients are more reliable than user location in CF mMIMO systems because user location and channel state information are not closely related.

\item Both the size of NN and dataset are small. This helps hugely reduce the time complexity for training the NN. Moreover, the proposed methods use unsupervised learning, which avoids a large burden of precomputation of optimal user power allocations.

\item We propose a general network that can be applied for max-min, max-product and max-sum-rate optimization with the same structure. Numerical results show that the proposed DL-based power control methods can achieve the same or better performance for three power control policies as the ones in previous works.

\end{itemize}


\section{System Model}
We consider a CF mMIMO system with $L$ randomly located single-antenna APs and $K$ single-antenna UEs following the uniform distribution, and all APs are connected to CPU via fronthauls. The channel coefficient of Rayleigh fading distribution between the user $k$ and the AP $l$ is modeled as $g_{kl} = \sqrt{\beta_{kl}}h_{kl}$, where $\beta_{kl}$ represents the LSF coefficient consisting of shadowing and pathloss, and $h_{kl}$ represents the small-scale fading coefficient, respectively. The uplink transmission of CF mMIMO system consists of pilot transmission, channel estimation, and uplink data transmission \cite{9110802,zheng2020efficient,liu2019tabu}.

\subsection{Pilot Transmission and Channel Estimation}
To estimate the uplink channel coefficients, all users transmit pilot sequences to all APs simultaneously at the beginning of uplink transmission in CF mMIMO systems. Let $\sqrt{\tau}\pmb{\phi}_k \in \mathbb{C}^{\tau \times 1}$ be the pilot sequence of $k$th UE, where $\tau$ is the length of pilot sequences, satisfying $\left\|\pmb{\phi}_k\right\|^2 = 1$. We assume that there is no pilot contamination because the number of UEs is smaller than the number of pilots. Thus the received signal at AP $l$ is
\begin{equation}
\textbf{y}_{p,l} = \sqrt{\tau\rho_p}\sum_{k=1}^{K} g_{kl}\pmb{\phi}_k + \textbf{w}_{p,l},
\end{equation}
where $\rho_p$ represents the normalized signal-to-noise ratio of pilot symbol, and $\textbf{w}_{p,l} \in \mathbb{C}^{\tau \times 1}$ is the additive noise at the $l$th AP and elements of $\textbf{w}_{p,l}$ are i.i.d $\mathcal{CN}$ (0, 1) random variables \cite{jin2019channel}. Then the estimation of channel coefficients $g_{kl}$ for $k$th UE is given by projecting $\textbf{y}_{p,l}$ onto $\pmb{\phi}_{k}^{H}$ as
\begin{equation}
\tilde y_{p,kl} = \sqrt{\tau\rho_p}g_{kl} + \sqrt{\tau\rho_p}\sum_{k' \neq k}^{K} g_{k'l}\pmb{\phi}_{k}^{H}\pmb{\phi}_{k'} + \pmb{\phi}_{k}^{H}\textbf{w}_{p,l}.
\end{equation}
And the linear minimum mean-squared error estimation of $g_{kl}$ is given by
\begin{equation}
\hat{g}_{kl} = \frac{\mathbb{E}\{\tilde y_{p,kl}^{*}g_{kl}\}}{\mathbb{E}\{\left|\tilde y_{p,kl}\right|^2\}}\tilde y_{p,kl} = c_{kl}\tilde y_{p,kl},
\end{equation}
where $c_{kl}$ is
\begin{equation}
c_{kl} = \frac{\sqrt{\tau\rho_p}\beta_{kl}}{\tau\rho_p\sum_{k'= 1}^{K}\beta_{k'l}\left|\pmb{\phi}_{k}^{H}\pmb{\phi}_{k'}\right|^2 + 1}.
\end{equation}

\subsection{Uplink Data Transmission}
For uplink data transmission, all users simultaneously send signals to all APs. Let $\rho$ and $\sqrt{\rho \eta_k}s_k$ represents the normalized uplink signal-to-noise ratio and the transmitted symbol of $k$th UE, satisfying $\mathbb{E}\{\left|s_k\right|^2\} = 1$. $\eta_k$ and $s_k$ are the power control coefficient and signal of $k$th UE, respectively. The received signal at $l$th AP is
\begin{equation}
x_l = \sqrt{\rho}\sum_{k=1}^K g_{kl}\sqrt{\eta_k}s_k + w_l,
\end{equation}
then the signals are combined at the CPU to detect $s_k$ \cite{rajapaksha2021deep}. With LSF coefficients known, the received signals at CPU is
\begin{equation}
r_{k} = \sqrt{\rho}\sum_{k'=1}^{K}\sum_{l=1}^{L} \sqrt{\eta_{k'}}\tilde g_{kl}^{*}g_{k'l}s_{k'} + \sum_{l=1}^{L}\tilde g_{kl}^{*}w_l.
\end{equation}
The uplink signal-to-interference-and noise ratio $\left(\text{SINR}\right)$ of $k$th UE is given by (\ref{SINR}) \cite{ngo2017cell}\cite{bjornson2019making}, where $\gamma_{kl} = \mathbb{E}\{|\hat{g}_{kl}|^2\} = \sqrt{\tau \rho_{p}}\beta_{kl} c_{kl}$. Thus the SE of $k$th UE is
\begin{equation}\label{SE}
\text{SE}_{k} = \left(1- \frac{\tau_p}{\tau_c}\right)\log_{2}\left(1+ \text{SINR}_k\right),
\end{equation}
where the factor $\left(1- \frac{\tau_p}{\tau_c}\right)$ in (\ref{SE}) is the fraction of channel uses that are dedicated for uplink data transmission \cite{bjornson2019making}.
\begin{figure*}[hb]
\hrule
\begin{align}\label{SINR}
\text{SINR}_k = \frac{\rho \eta_k \left(\sum\limits_{l=1}^{L}\gamma_{kl} \right)^2}{ \rho \sum\limits_{k' \neq k}^{K}\eta_{k'}\left(\sum\limits_{l=1}^{L}\gamma_{kl}\frac{\beta_{k'l}}{\beta_{kl}}\right)^2 \left|\pmb{\phi}_{k}^{H}\pmb{\phi}_{k'}\right|^2 + \rho\sum\limits_{k' = 1}^{K}\eta_{k'}\sum\limits_{l=1}^{L}\gamma_{kl}\beta_{k'l}+\sum\limits_{l=1}^{L}\gamma_{kl}}.
\end{align}
\end{figure*}

\section{Max-Min, Max-Product and Max-Sum-Rate Power Control Optimization}
The optimization-based power control methods for max-min, max-sum-rate and max-product power control optimization are introduced as follows.
The uplink max-min power control scheme is
\begin{equation}
\begin{aligned}
&\mathop{\rm{max}}\limits_{\eta_k}    \mathop{\rm{min}}\limits_{k=1,2,\ldots,K} \text{SE}_k,\\
&s.t.  \quad 0\leq \eta_k \leq 1, \quad k=1,2,\ldots,K.
\end{aligned}
\end{equation}
As proposed in \cite{ngo2017cell,zhang2018performance}, max-min optimization aims to provide uniform service to all UEs and is programmed to an algorithm using bisection search to solve the optimization problem mentioned above.

Max-sum-rate optimization aims to maximize the sum-rate of all UEs. The max-sum-rate scheme can be achieved by solving the optimization problem as follows \cite{bai2020sum}
\begin{equation}
\begin{aligned}
&\mathop{\rm{max}}\limits_{\eta_k}   \sum\limits_{k=1}^K \text{SE}_k,\\
&s.t.  \quad 0\leq \eta_k \leq 1, \quad k=1,2,\ldots,K.
\end{aligned}
\end{equation}

Max-product optimization is proposed to maximize the product of SEs or SINRs directly. We refer to the model in \cite{sanguinetti2018deep} and formulate the problem as follows
\begin{equation}
\begin{aligned}
&\mathop{\rm{max}}\limits_{\eta_k}   \prod\limits_{k=1}^K \text{SINR}_k,\\
&s.t.  \quad 0\leq \eta_k \leq 1, \quad k=1,2,\ldots,K.
\end{aligned}
\end{equation}

However, such an aforementioned analytical method is time-consuming and not scalable enough, which makes it unfeasible for large systems. In the following section, we propose a DL-based approach to learn the optimal power control coefficients for max-min, max-product and max-sum-rate optimization.

\section{DL-Based Power Control}
DL has attracted much interest for the solution of optimal power control in CF mMIMO systems. Most of the aforementioned works are based on supervised learning \cite{sanguinetti2018deep, d2019uplink}, which requires intensive training data for every system status that means high computational complexity. Compared with supervised learning, a time-consuming method, unsupervised learning (UL) does not require optimal power control outputs for training data, therefore there is no need to solve optimization problems repeatedly. In contrast, UL avoids precomputation of optimal power control coefficients and takes LSF coefficients as input to directly minimize the given loss function, which is related to the target power control policy. The scalability of UL is superior to that of convex solvers \cite{nikbakht2020unsupervised}, which is essential when $L$ and $K$ are large. Because of its flexibility, UL is easier to implement and can achieve better performance.

Due to the reason that the power of UEs can be optimized based on only the LSF coefficients instead of taking small-scale fading coefficients into account \cite{ghazanfari2020enhanced}, in particular, we form an aggregated LSF coefficient for each user $k$ as follows:

\begin{equation}
B_{k} = \sum_{l=1}^{L} \beta_{kl},
\end{equation}
where $B_k$ is the summation of LSF coefficients of $k$th UE with different APs. As the numbers of UEs and APs increase in the CF mMIMO systems, the input $\left(\beta_k\right)$ of NNs increase with it, making the training process complicated and time-consuming. It can be verified that using $B_k$ as inputs does not have much performance loss and can highly reduce computational complexity. The procedure of the proposed method is shown in Fig. 1, where the input of NN is $B_k$ instead of LSF coefficients given the above reasons.

As $B_k$ is the input of rectified linear unit (ReLU) activation function, to fully learn and outperform the map between power control coefficients $p_k$ for all UEs, the input layer is followed by two hidden layers using ReLU as the activation function. The sigmoid activation function is used to get the power control coefficients as output.

\begin{figure}[t]
\centering 
\includegraphics[width=0.40\textwidth]{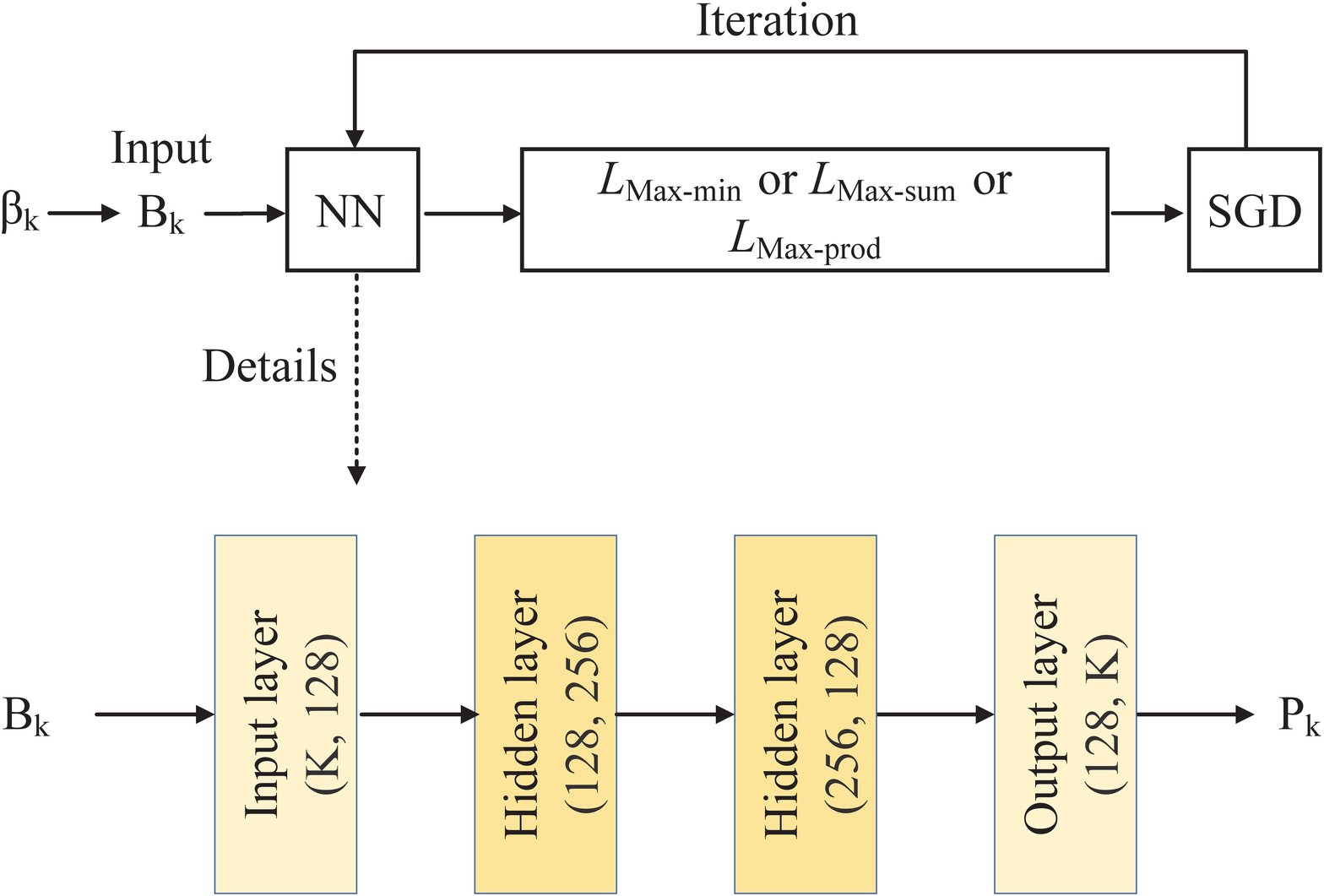} 
\caption{DL-based power control method for uplink CF mMIMO system.} 
\label{Fig.network} 
\end{figure}

\subsection{DL-Based Max-Min}
Max-min optimization aims to provide uniform service to all UEs and is widely used in CF mMIMO systems. Compared with existing works \cite{ngo2017cell, rajapaksha2021deep}, we improve the UL method, especially the loss function and NNs.

Letting $S(\cdot)$ represent a sigmoid function, the loss function of max-min optimization is designed as
\begin{equation}\label{maxmin}
L_{\text{Max-min}} = \sum_{k=1}^{K} S\left(\frac{0.3}{\text{SINR}_k}\right) - \alpha \times \text{SINR}_{\rm min},
\end{equation}
where $\alpha$ is a hyperparameter that controls the proportion of $\text{SINR}_{\rm{min}}$. To make the gradient descent of the training process more stable, the robustness of the NN needs to be improved. The main challenges of the gradient descent process are as follows:
\begin{itemize}
\item Exploding gradient problem. When using multiple layers and loss functions with large gradient, the NN parameters will be greatly updated, causing the NN to be unstable.
\item Vanishing gradient problem. Using the loss function with the small gradient will cause slowly update of NN parameters. Besides, the gradient descent slows down as the number of NN layers and size of NN increase.
\end{itemize}
This problem was for instance noticed in \cite{rajapaksha2021deep}, where the following loss function was used:
\begin{equation}\label{maxmin-other}
L =  - \alpha \times \text{SINR}_{\rm min}.
\end{equation}
Iterating the gradient descent process with only $\text{SINR}_{\rm min}$ probably slows down the gradient descent. Thus, we use the loss function that contains sigmoid function. Compared with $\left(\ref{maxmin-other}\right)$, the proposed $L_\text{{Max-min}}$ performs better and reduces computational complexity. And it is more stable than other loss functions, especially loss functions that contain scores $\left(\text{e.g., } L = \sum_{k=1}^{K} e^{\frac{\alpha}{\beta + \text{SINR}_k}}\right)$. The detailed results will be analyzed in the subsection of the numerical results.

\subsection{DL-Based Max-Sum-Rate}
Max-sum-rate optimization aims to maximize the summation of SE, like $\sum_{k=1}^K \text{SE}_k$, and the loss function of max-sum-rate optimization can be posed as the minimization over power coefficients $\eta_k$ of the cost

\begin{equation}\label{maxsum}
\begin{aligned}
L_{\text{Max-sum-rate}} & = - \sum_{k=1}^{K} \frac{1}{\mu_k}\log_2 \left(1+\text{SINR}_k\right) \\
\end{aligned}
\end{equation}
where $\mu_k$ controls the proportion of $\text{SE}_k$. Different from max-min and max-product optimization, max-sum-rate optimization is more difficult to learn because of the small gradient. Hence, more epochs and training data are required by NN.

Owing to the efficient gradient descent and less preparation/training time, it is recommended to learn max-product optimization by DL, as illustrated in the next section.

\subsection{DL-Based Max-Product}
Max-product optimization aims to maximize the production of SE or SINR, like $\prod_{k=1}^{K} \text{SE}_k$. To make the gradient descent more available and stable, we use a logarithmic function instead of product, which avoids the vanishing gradient problem.
\begin{equation}\label{maxprod}
\begin{aligned}
L_{\text{Max-prod}} & = - \sum_{k = 1}^{K} \gamma_k\log_2 \log_2 \left(1+\text{SINR}_k\right) \\
\end{aligned}
\end{equation}
where $\gamma_k$ represents the proportion of $k$th UE.

Compared with DL-based max-sum-rate optimization, the loss function (\ref{maxprod}) contains logarithm, which works better in gradient descent. Hence DL-based max-product with the loss function (\ref{maxprod}) is more easily trained and converges faster.

\section{Simulations and Numerical Results}
In this section, hyperparameters of DL are presented. Using these parameters, we derive the DL method of three different power control policies and evaluate its performance in comparison with optimization-based methods.

\subsection{Simulation Setups}
We consider a CF mMIMO system in a square area of size 1 $\times$ 1 $\text{km}^2$, which is wrapped around at the edges to avoid boundary effects and to simulate a CF mMIMO network with an infinite area. The LSF coefficient $\beta_{kl}$ between $k$th UE and $l$th AP is given as
$\beta_{kl} = \text{PL}_{kl}10^{\frac{\sigma_{\text{sh}}z_{kl}}{10}}$ \cite{ngo2017cell}, where $\text{PL}_{kl}$ is the pathloss calculated by the three-slope model \cite{ngo2017cell}, and the shadow fading with standard derivation $\sigma_{\text{sh}}$, $z_{kl} \sim \mathcal{N}\left(0,1\right)$ is $10^{\frac{\sigma_{\text{sh}}z_{kl}}{10}}$. $\tau = 20$ is set in the simulations. We refer to Table \uppercase\expandafter{\romannumeral1} in \cite{ngo2017cell} for detailed parameters of the CF mMIMO system.

The original problems have been solved as explained in \cite{ngo2017cell, sanguinetti2018deep}. When it comes to DL, we completed the study by PyTorch including training and testing. The training and testing works are done with an Nvidia GeForce GTX 1660Ti $(6\rm{GB})$ Graphics Processing Unit.

To make gradient descent more available, stochastic gradient descent (SGD) optimizer is applied, which is better than Adam in this simulation environment. The training data for max-min, max-product and max-sum-rate optimizations consists of 10000 different samples of CF mMIMO systems, respectively. Specifically, $\alpha = 1$, $\mu_k=5, \forall k$ and $\gamma_k = 1, \forall k$ are set for the loss function (\ref{maxmin}), (\ref{maxsum}) and (\ref{maxprod}) in this paper. The network for max-min, max-sum-rate and max-product optimization is trained with a start learning rate of 0.3, 1 and 0.03, respectively. 300 updates of the NN take place, and to reduce random fluctuations in the training process, the learning rate is adjusted when reaching 150 epochs during training.

\subsection{Numerical Results}
\begin{figure}[t]
\centering 
\includegraphics[width=0.40\textwidth]{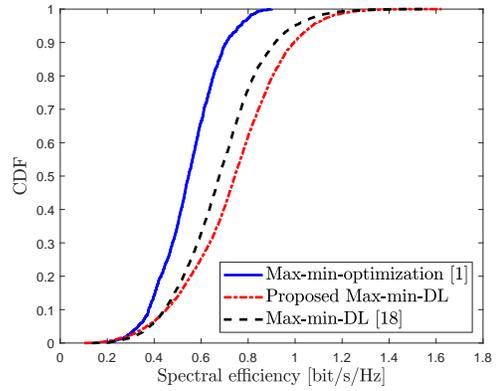} 
\caption{CDF of the SE for max-min optimization with different methods ($K = 8$, $L = 20$).} 
\label{Fig.maxmin} 
\end{figure}
The performance of the proposed method is analyzed below.
\subsubsection{Max-Min}

Fig. \ref{Fig.maxmin} shows cumulative distribution function (CDF) of the SE of randomly located UEs with max-min optimization using three different methods, consisting of optimization-based method, proposed DL-based method and DL-based method in \cite{rajapaksha2021deep}. The DL method for max-min optimization proposed in this study outperforms the optimization method and provides better overall performance compared with loss function (\ref{maxmin-other}).

Compared with the loss function which uses the mathematical constant $e$ and score as an exponent $\left(\text{e.g., } L = \sum_{k=1}^{K} e^{\frac{\alpha}{\beta + \text{SINR}_k}}\right)$, the loss function (\ref{maxmin}) avoids the vanishing gradient problem, thus providing a more stable training process. The performance of loss function (\ref{maxmin}) does not rely much on fixed hyperparameters, hence the numerator can be replaced by a smaller coefficient and without having much impact on performance.

\begin{figure}[t]
\centering 
\includegraphics[width=0.40\textwidth]{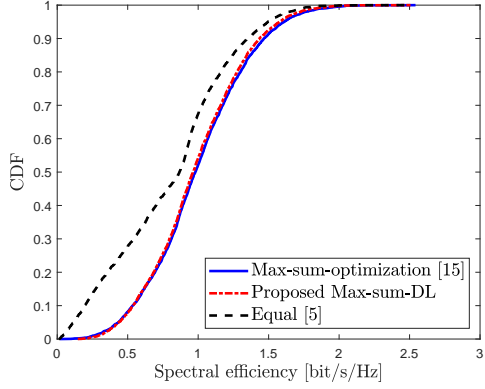} 
\caption{CDF of the SE for max-sum-rate optimization with different methods ($K = 20$, $L = 50$).} 
\label{Fig.maxsum} 
\end{figure}

\begin{figure}[t]
\centering 
\includegraphics[width=0.40\textwidth]{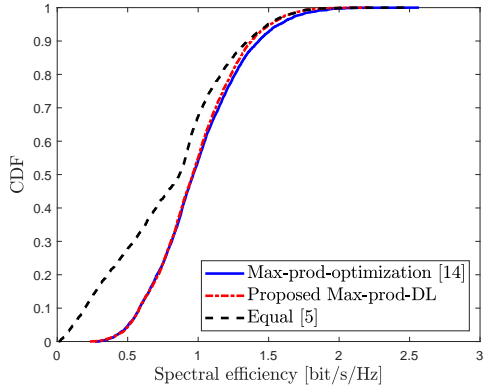} 
\caption{CDF of the SE for max-product optimization with different methods ($K = 20$, $L = 50$).} 
\label{Fig.maxprod} 
\end{figure}

\subsubsection{Max-Sum-Rate}
The performance of optimization-based, DL-based and equal power control methods for max-sum-rate optimization is analyzed in Fig. \ref{Fig.maxsum} with $K = 20$ and $L = 50$. It can be seen that both DL-based and optimization-based method outperforms equal power control that allocates all UEs with the maximum power, and the curve of the proposed DL-based method fits well compared with the curve of optimization-based method. DL-based method proposed in this paper takes advantage of unsupervised learning, which does not require optimal power allocations to be known and hence has lower complexity.

\subsubsection{Max-Product}
Fig. \ref{Fig.maxprod} portrays the performance of optimization-based, DL-based and equal power control methods for max-product optimization. There is a big gap between the equal power control method and the two max-product methods. The curve of DL-based method matches closely its optimization-based method. When it comes to the comparison of max-sum-rate and max-product optimization, there are two physical insights as follows:
\begin{itemize}
\item Compared with max-sum-rate optimization, max-product optimization has a better 95\%-likely performance.
\item Max-product optimization is easier to learn by DL. Compared with max-product, max-sum-rate optimization has a gap between optimization and DL performance when training data is insufficient (e.g., 4000 samples) and thus requires a larger dataset. This is because the loss function (\ref{maxsum}) has a smaller gradient.
\end{itemize}
Given the above reasons, it is highly recommended to use max-product optimization instead of max-sum-rate optimization.

\begin{figure}[t]
\centering 
\includegraphics[width=0.40\textwidth]{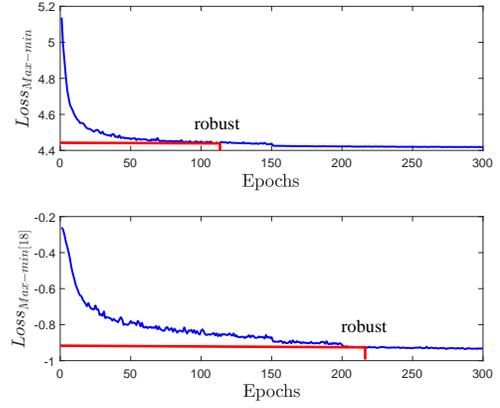} 
\caption{Learning curves for DL-based max-min optimization.} 
\label{Fig.loss-maxmin} 
\end{figure}

\subsubsection{Loss Function}
The learning curves for different DL-based power control methods for max-min optimization are shown in Fig. \ref{Fig.loss-maxmin}, where the subgraph at the top of the figure is the learning curve of the loss function (\ref{maxmin}) and the bottom is the loss function (\ref{maxmin-other}). The DL-based power control method proposed in this paper converges in a short time, achieving a robust performance. The convergence position is marked in the figure, at about 120 epochs, while the method in \cite{rajapaksha2021deep} with the loss function (\ref{maxmin-other}) has a longer convergence time at more than 200 epochs.

\begin{figure}[t]
\centering 
\includegraphics[width=0.40\textwidth]{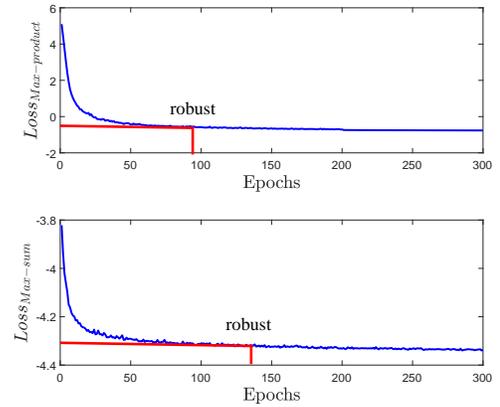} 
\caption{Learning curves for DL-based max-product and max-sum-rate optimizations.} 
\label{Fig.loss-maxsum/prod} 
\end{figure}

Figure \ref{Fig.loss-maxsum/prod} portrays the learning curves for DL-based power control for max-product and max-sum-rate optimizations. It is seen that both the two curves converge in a short time and the gradient drops smoothly in Fig. \ref{Fig.loss-maxsum/prod}. The convergence position is marked in the figure. The DL-based power control for max-sum-rate optimization is robust at less than 140 epochs, while the DL-based power control for max-product optimization is robust at around 100 epochs, which is faster than max-min and max-sum-rate optimization.

\begin{table}[t]
\centering
\caption{Computational Complexity of the Proposed DL-Based Method and Optimization-Based Method.}
\begin{tabular}{|c|c|c|}
  \hline
  \hline
    & \tabincell{c}{ Optimization-\\ Based Method [s]}& \tabincell{c}{DL-Based \\Method [s]}\\
  \hline
  Max-min  & 1.43  & 9.21 $\times$ $10^{-2}$  \\
  \hline
  Max-product  & 1.14  & 9.24 $\times$ $10^{-2}$  \\
  \hline
  Max-sum-rate  & 1.26 & 9.29 $\times$ $10^{-2}$  \\
  \hline
  \hline
\end{tabular}\label{table1}\vspace{-0.5em}
\end{table}

\subsubsection{Computational Complexity}
Table \ref{table1} shows the comparison of DL-based and optimization-based power control method. We consider CF mMIMO systems with $K = 8$, $L = 20$, $K = 20$, $L = 50$ and $K = 20$, $L = 50$ for max-min, max-product and max-sum-rate optimizations, respectively. The cost of optimization-based method is an average time of testing with a dataset consisting of 1 sample, in contrast, the cost of DL-based method is an average time of testing with a dataset consisting of 200 samples. Once trained, the NN can output power coefficients in 1 second. Due to similar DNN models, the time cost of three DL-based methods are almost the same, and they are about 15, 12 and 13 times faster than the optimization-based method for max-min, max-product and max-sum-rate optimizations, respectively. The training time is not a primary problem, due to the reason that there is no need to retrain the NN when the number of users is fixed.

Instead of using $B_k$, using $\beta_k$ as input will significantly increase the number of nodes in the neural network and increase the time cost of training, while this does not improve performance much. Hence it is highly recommended to use $B_k$ directly as input instead of $\beta_k$, which effectively reduces the size of the network.

\section{Conclusion}
In this study, a DL-based power control method has been proposed for max-min, max-product and max-sum-rate optimizations to learn the map between LSF coefficients and power coefficients in uplink CF mMIMO systems. The proposed method for max-min optimization improves overall performance and hugely reduces computational complexity compared with the supervised learning-based and optimization-based power control method. The method for max-product and max-sum-rate optimizations results in close SE curves compared with the optimization-based method. All three methods take advantage of no preparation for optimal power allocations and the small size of NN. It is important to note that the proposed DL-based power control method converges with a limited training time and has low computational complexity. Finally, we will consider deep reinforcement learning-based power control in future works.

\bibliographystyle{IEEEtran}
\bibliography{ref}

\end{document}